# Magnetic monopoles and Lorentz force

**F. Moulin**

Département de Physique, Ecole Normale Supérieure de Cachan, 94235 Cachan, France
**e-mail:** moulin@physique.ens-cachan.fr

**Abstract:** The formulation of a generalized classical electromagnetism that includes both electric and magnetic charges, is explored in the framework of two potential approach. It is shown that it is possible to write an action integral from which one can derive, by least action principle, the symmetrized set of Maxwell's equations, but also the Lorentz force law by employing the energy-momentum tensor conservation.

## 1. Introduction

Dirac [1] was the first to suggest the possibility of a particle that carries magnetic charge. At the present time there is no experimental evidence for the existence of magnetic charges or monopoles. Dirac's interresting argument, is that the mere existence of one magnetic monopole in the universe would offer an explanation of the discrete nature of electric charge. He introduced a singular vector potential and found that magnetic monopoles would have strings attached to them. N.Cabibbo and E.Ferrari [2] have shown that if one introduce a second four-potential it is possible to eliminate the Dirac string. Some authors [3,4,5], who continued the study of these ideas, showed that this two potentials original approach leads to a symmetrized Maxwell theory. Let us suppose here that exists particles that carries magnetic as well electric charges. It is shown, in this article, that it is possible to write an action integral from which one can derive, by least action principle, the symmetrized set of Maxwell's equations but also the Lorentz force law for both electrically and magnetically charged particles. For the first time in this case, the derivation of proper equation of motion is made possible here, by employing the energy-momentum tensor conservation in a curvilinear coordinates. The symmetrized Lorentz force density and a classical model of magnetic monopoles, are then obtained here directly from a Lagrangian fonction.

## 2. Fields, potentials and gauge invariance

We first define two four-vector potentials $A^i$ and $C^i$, where $A^i = (V, A)$ is the usual electromagnetic vector potential, and $C^i = (V', C)$ the "magnetoelectric" vector potential.

The equations governing the evolution of the system shall be invariant under usual gauge transformations :

(1) $\qquad A^{*i} = A^i - \partial^i \psi$

(2) $\qquad C^{*i} = C^i - \partial^i \psi'$

where $\psi(r,t)$ and $\psi'(r,t)$ are any scalar fields. We employ the convenient notation :

$$\partial^i = \frac{\partial}{\partial x_i} = (\frac{1}{c}\frac{\partial}{\partial t}, -\nabla), \quad \partial_i = \frac{\partial}{\partial x^i} = (\frac{1}{c}\frac{\partial}{\partial t}, \nabla) \text{ with } x_i = (ct, -r) \text{ and } x^i = (ct, r)$$

A total electromagnetic field tensor, $F^{ik}$, is invariant under gauge transformations when it can be writen, with $A^i$ and $C^i$, by [2,3] :

(3) $$F^{ik} = \delta^{ik}_{pq} \partial^p A^q - \varepsilon^{ikpq} \partial_p C_q$$

where $\varepsilon^{ikpq}$ is the completely antisymmetric unit tensor of the fourth rank ($\varepsilon^{0123} = -\varepsilon_{0123} = +1$ is the same convention found in ref [6,7,8] but opposite in ref [3] )

The tensor $\delta^{ij}_{pq}$ is a new and convenient notation for calculations, defined as a combination of Kronecker symbols in four-dimensional space : $\delta^{ij}_{pq} = \delta^i_p \delta^j_q - \delta^i_q \delta^j_p$ and then $\delta^{ik}_{pq} \partial^p A^q = \partial^i A^k - \partial^k A^i$.

Making use of relations (1), (2) and (3), we obtain :

$$F^{*ik} = \delta^{ik}_{pq} \partial^p A^{*q} - \varepsilon^{ikpq} \partial_p C^*_q = F^{ik}$$

where it is easy to demonstrate that : $\varepsilon^{ikpq} \partial_p \partial_q \psi' = 0$ and $\delta^{ik}_{pq} \partial^p \partial^q \psi = 0$

$F^{ik}$ is then invariant under gauge transformation.

The dual tensor $G^{ik}$ of the electromagnetic four rank tensor $F^{ik}$, is also invariant under gauge transformations. $G^{ik}$ is defined by :

(4) $$G^{ik} = \frac{1}{2} \varepsilon^{ikpq} F_{pq}$$

With (3) and (4) we can write :

(5) $$G^{ik} = \delta^{ik}_{pq} \partial^p C^q + \varepsilon^{ikpq} \partial_p A_q$$

where we employ the relations:

(6) $$\varepsilon^{ijpq} \varepsilon_{klpq} = -2\delta^{ij}_{kl} \quad \text{and} \quad \varepsilon^{ijpq} \delta^{kl}_{pq} = 2\varepsilon^{ijkl}$$

Making use of relations (1), (2) and (5), we obtain :

$$G^{*ik} = \delta^{ik}_{pq} \partial^p C^{*q} - \varepsilon^{ikpq} \partial_p A^*_q = G^{ik}$$

$G^{ik}$ is then also invariant under gauge transformation.

The field tensor $F^{ik}$ is, in matrix standard form [7,8] :

(7) $$F^{ik} = \begin{pmatrix} 0 & -E_x & -E_y & -E_z \\ E_x & 0 & -B_z & B_y \\ E_y & B_z & 0 & -B_x \\ E_z & -B_y & B_x & 0 \end{pmatrix}$$

The elements of $F_{ik}$ are obtained from $F^{ik}$ by putting $\mathbf{E} \to -\mathbf{E}$

The elements of the dual field tensor $G^{ik}$ are obtained from $F^{ik}$ with (4) and (7) :

$$(8) \quad G^{ik} = \begin{pmatrix} 0 & -B_x & -B_y & -B_z \\ B_x & 0 & E_z & -E_y \\ B_y & -E_z & 0 & E_x \\ B_z & E_y & -E_x & 0 \end{pmatrix}$$

The elements of $G_{ik}$ are obtained from $G^{ik}$ by putting $\boldsymbol{B} \to -\boldsymbol{B}$

The reader can check that the vectorial relations betwen the fields (*E, B)* and the potentials (*V,A*), *(V',C)* are given by :

$$(9) \quad \begin{cases} \boldsymbol{E} = -\nabla \times \boldsymbol{C} - \dfrac{1}{c}\dfrac{\partial \boldsymbol{A}}{\partial t} - \nabla V \\ \boldsymbol{B} = \nabla \times \boldsymbol{A} - \dfrac{1}{c}\dfrac{\partial \boldsymbol{C}}{\partial t} - \nabla V' \end{cases}$$

## 3. Maxwell's equations from least action principle

*a. Euler-Lagrange equations*

The Maxwell's equations can be derived from an action integral with a least action principle. The action formulation contains terms with the two four-vector potentials $A_k$ and $C_k$ and with their derivate $\partial_i A_k$ and $\partial_i C_k$ . The action is associated with a Lagrangian density $\Lambda$ by the relation :

$$(10) \quad S = \frac{1}{c}\int \Lambda(A_k, C_k, \partial_i A_k, \partial_i C_k)\, d\Omega$$

The Euler-Lagrange equations follow from the stationary property of the action integral with respect to variations $\delta A_k$ and $\delta C_k$ :

$$(11) \quad \partial_i \left(\frac{\partial \Lambda}{\partial(\partial_i A_k)}\right) - \frac{\partial \Lambda}{\partial A_k} = 0$$

$$(12) \quad \partial_i \left(\frac{\partial \Lambda}{\partial(\partial_i C_k)}\right) - \frac{\partial \Lambda}{\partial C_k} = 0$$

*b. Lagrangian density*

We now examine the lagrangian description of the field in interaction with specified external sources. The interaction terms in $\Lambda$ involves magnetic current density $k^i = (\rho_m c, \boldsymbol{j}_m)$ and electric current density $j^i = (\rho_e c, \boldsymbol{j}_e)$ associated with their respective vector potential $C^i$ and $A^i$.

For a standard electromagnetic theory, with no monopoles, the usual and well known Lagrangian $\Lambda_e$, is given by [6,7,8] : $\Lambda_e = -\frac{1}{c}j^i A_i - \frac{1}{16\pi}F^{ik}F_{ik}$ where $F^{ik}F_{ik}$ is a Lorentz-invaraint quadratic form associated with the electromagnetic field $F^{ik} = \delta^{ik}_{pq}\partial^p A^q$.

With the presence of magnetic sources, we believe that a $\frac{1}{c}k^i C_i$ interaction term is simply introduce inside the lagrangian function. We also believe that the free fields from electric or magnetic sources are not distinguishable from each other.

$$(13) \quad \Lambda = \frac{1}{c}k^i C_i - \frac{1}{c}j^i A_i - \frac{1}{16\pi}F^{ik}F_{ik}$$

where $F^{ik} = \delta^{ik}_{pq}\partial^p A^q - \varepsilon^{ikpq}\partial_p C_q$ is now the total electromagnetic field.

(I use here Gaussian units which are particularly convenient because electric and magnetic charges are expressed in the same units)

*c. Maxwell's equations*

In order to use the Euler-Lagrange equations (11)(12) and with (13) and (4), we can derive the symmetrized Maxwell's equations in a covariant form :

$$(14) \quad \partial_i F^{il} = \frac{4\pi}{c}j^l$$

$$(15) \quad \partial_i G^{il} = \frac{4\pi}{c}k^l$$

In vectorial notations, the set of Maxwell's equations is :

$$(16) \quad \begin{cases} \nabla . \boldsymbol{E} = 4\pi\rho_e & \nabla \times \boldsymbol{B} = \frac{4\pi}{c}\boldsymbol{j}_e + \frac{1}{c}\frac{\partial \boldsymbol{E}}{\partial t} \\ \nabla . \boldsymbol{B} = 4\pi\rho_m & -\nabla \times \boldsymbol{E} = \frac{4\pi}{c}\boldsymbol{j}_m + \frac{1}{c}\frac{\partial \boldsymbol{B}}{\partial t} \end{cases}$$

The conservation of the source current density can be obtained by taking the four-divergence of both side of Maxwell's equations (14) and (15) : $\partial_l \partial_i F^{il} = \frac{4\pi}{c}\partial_l j^l$ and $\partial_l \partial_i G^{il} = \frac{4\pi}{c}\partial_l k^l$. $F^{il}$ and $G^{il}$ are antisymmetric tensors, then $\partial_l \partial_i F^{il} = 0$ and $\partial_l \partial_i G^{il} = 0$. The continuity equations for electric and magnetic current densities are then given by :

(17) $\quad\quad\quad \partial_l j^l = 0$

(18) $\quad\quad\quad \partial_l k^l = 0$

*d. Laplace equations*

Substituting for $F^{il}$ and $G^{il}$ their expressions in term of the potentials (3) (5), the Maxwell's equations (14) and (15) become : $\partial_i \partial^i A^l - \partial_i \partial^l A^i = \frac{4\pi}{c} j^l$ and $\partial_i \partial^i C^l - \partial_i \partial^l C^i = \frac{4\pi}{c} k^l$.

We impose on the potentials $A^i$ and $C^i$ the usual Lorentz conditions :

(19) $\quad\quad\quad \partial_i A^i = 0$

(20) $\quad\quad\quad \partial_i C^i = 0$

and then the Laplace equations are given by :

(21) $\quad\quad\quad \partial_i \partial^i A^l = \frac{4\pi}{c} j^l$

(22) $\quad\quad\quad \partial_i \partial^i C^l = \frac{4\pi}{c} k^l$

$\partial_i \partial^i$ is the four-dimensional Laplacian operator or d'Alembert operator. The solutions of this type of equations are found, for example, in ref [6,7].

## 4. Energy-momentum tensor and symmetrized Lorentz force

Variation of action integral, with respect to the particle coordinates, don't lead directly to the correct equation of motion for electrically and magnetically charged particles.

To obtain the correct symmetrized Lorentz force density, we shall use, for the first time in this case, the conservation of the total energy momentum tensor $T^{ik}$ in curvilinear coordinates :

(23) $\quad\quad\quad D_i T^{ik} = 0$

where $D_i T^{ik}$ is the usual covariante derivates of the total energy-momentum tensor.

This conservation law is a direct consequence of the Einstein equation of the general relativity : $R_{ij} - \frac{1}{2} g_{ij} R = \chi T_{ij}$ where $R_{ij}$ is the Ricci tensor, $R$ the scalar curvature and $\chi$ a constant ( see for example ref [6], [8] or [9] ).

*a*. *Calculation of the energy-momentum tensor*

In curvilinear coordinates the general action integral must be writen in the form :

$$(24) \quad S = \frac{1}{c}\int \sqrt{-g}\,\Lambda(g_{ik},\partial_l g_{ik})\,d\Omega$$

Where the quantities $g_{ik}$ represent the space-time metric, which determines all the geometric properties in each system of curvilinear coordinates. $g$ is the determinant formed from the quantities $g_{ik}$ ( $g$ is always negative for a real space-time, $g=-1$ in Galilean coordinates ).
The variation of the action, with respect to the space-time metric, leads to the definition of the general energy-momentum tensor $T^{ik}$ [6] :

$$(25) \quad T^{ik} = \frac{2}{\sqrt{-g}}\left[\frac{\partial(\sqrt{-g}\,\Lambda)}{\partial g_{ik}} - \partial_l\left(\frac{\partial(\sqrt{-g}\,\Lambda)}{\partial(\partial_l g_{ik})}\right)\right]$$

This formula is convenient for calculating the energy-momentum tensor not only in the case of the presence of gravitational field, but also in its absence, or, more interresting, in the presence of a total electromagnetic field. The curvilinear coordinates occurs formally as an intermediary step in the calculation of $T^{ik}$.
For the free electromagnetic field, in a source free region with no gravitational field, the Lagrangian
$\Lambda$ is given by (13) :

$$(26) \quad \Lambda = -\frac{1}{16\pi}F_{pq}F^{pq} = -\frac{1}{16\pi}g_{pr}g_{qs}F^{rs}F^{pq}$$

$\Lambda$ is not dependant of $\partial_l g_{ik}$, then the second term in (25) vanish. Using relation (25), we find the symmetric tensor expression for electromagnetic field :

$$(27) \quad T^{(f)ik} = \frac{1}{4\pi}\left[-F^{il}F^{k}{}_{l} + \frac{1}{4}g^{ik}F^{pq}F_{pq}\right]$$

where we use relations : $\dfrac{\partial g_{lj}}{\partial g_{ik}} = \delta_l^i \delta_j^k$ and $\dfrac{\partial \sqrt{-g}}{\partial g_{ik}} = -\frac{1}{2}\sqrt{-g}\,g^{ik}$

Or in the form of the mixed tensor $T^{(f)i}{}_k$ :

$$(28) \quad \begin{aligned} T^{(f)i}{}_k &= T^{(f)ij}g_{jk} = \frac{1}{4\pi}\left[-F^{il}F^{j}{}_{l}g_{jk} + \frac{1}{4}g^{ij}g_{jk}F^{pq}F_{pq}\right] \\ &= \frac{1}{4\pi}\left[F^{li}F_{kl} + \frac{1}{4}\delta_k^i F^{pq}F_{pq}\right] \end{aligned}$$

We note that $T^{(f)ik}$ is a symmetric tensor $T^{(f)ik} = T^{(f)ki}$, and the sum of the diagonal term is zero $T^{(f)i}{}_i = 0$.

We now introduce the $G^{ik}$ tensor inside the energy-momentum tensor $T^{(f)i}{}_k$. We demonstrate first the following relation :

$$
\begin{aligned}
G^{li}G_{kl} &= (\frac{1}{2}\varepsilon^{lipq}F_{pq})(\frac{1}{2}\varepsilon_{klrs}F^{rs}) \\
&= -\frac{1}{4}\varepsilon^{pqil}\varepsilon_{rskl}F_{pq}F^{rs} \\
&= \frac{1}{4}\delta^{pqi}_{rsk}F_{pq}F^{rs} \\
&= F^{li}F_{kl} + \frac{1}{2}\delta^i_k F^{pq}F_{pq}
\end{aligned}
\tag{29}
$$

where the tensor $\delta^{pqi}_{rsk}$ is defined by a combination of products of components of the unit tensor $\delta^i_j$ in four-dimensional space [6,9] :

$$
\delta^{pqi}_{rsk} = -\varepsilon^{pqil}\varepsilon_{rskl} = \begin{vmatrix} \delta^p_r & \delta^p_s & \delta^p_k \\ \delta^q_r & \delta^q_s & \delta^q_k \\ \delta^i_r & \delta^i_s & \delta^i_k \end{vmatrix} = \delta^p_r\delta^q_s\delta^i_k + \delta^p_s\delta^q_k\delta^i_r + \delta^p_k\delta^q_r\delta^i_s - \delta^p_r\delta^q_k\delta^i_s - \delta^p_s\delta^q_r\delta^i_k - \delta^p_k\delta^q_s\delta^i_r
$$

The $T^{(f)i}{}_k$ tensor can be write finally in a useful new symmetrized form by using the identity (29) :

$$
T^{(f)i}{}_k = \frac{1}{8\pi}\left[F^{li}F_{kl} + G^{li}G_{kl}\right] \tag{30}
$$

*b . The Lorentz force density*

In the presence of a gravitational field, the symmetrized Lorentz force density is obtained from the conservation of $T^i{}_k$ in curvilinear coordinates $D_i T^i{}_k = 0$ (23). As already mentioned, the curvilinear coordinates occur formally as an intermediary step in the calculation of $T^{ik}$. The relation (30) is also valid in the absence of a gravitational field. The covarariante derivates $D_i$ then becomes $\partial_i$ in equations.

The total energy momentum tensor $T^i{}_k$ of the system, is defined as the sum of the energies and momenta of the fields and particles. In other words, we shall verify the conservation equation :

(31) $\quad \partial_i T^i{}_k = \partial_i (T^{(f)i}{}_k + T^{(p)i}{}_k ) = 0$

where $T^{(p)i}{}_k = \rho_0 u^i u_k$ [6,8] is the energy-momentum tensor of the particles. $u^i = \dfrac{dx^i}{dt_0} = (\gamma c, \gamma v)$ is the particle's four-velocity, $\rho_0$ the proper mass density and $t_0$ the proper time.

Differentiating the expression of $T^{(p)i}{}_k$ gives :

(32)
$$\begin{aligned}
\partial_i T^{(p)i}{}_k &= u_k \partial_i (\rho_0 u^i) + \rho_0 u^i \partial_i u_k \\
&= \rho_0 \frac{dx^i}{dt_0} \frac{\partial u_k}{\partial x_i} \\
&= \rho_0 \frac{du_k}{dt_0} \\
&= f_k
\end{aligned}$$

where $\partial_i (\rho_0 u^i) = 0$ is the mass current law conservation, as analogous to the charge current conservation. $f_k$ is defined as the four-vector force density :

(33) $\quad f_k = \partial_i T^{(p)i}{}_k = -\partial_i T^{(f)i}{}_k$

Making use of the covariant Maxwell equations (14) (15), with the relation (30), and after some manipulations, we find as solution for $\partial_i T^{(f)i}{}_k$ :

(34)
$$\begin{aligned}
\partial_i T^{(f)i}{}_k &= \frac{1}{8\pi} \left[ F^{li} \partial_i F_{kl} + F_{kl} \partial_i F^{li} + G^{li} \partial_i G_{kl} + G_{kl} \partial_i G^{li} \right] \\
&= \frac{1}{4\pi} \left[ F_{kl} \partial_i F^{li} + G_{kl} \partial_i G^{li} \right] \\
&= -\frac{1}{c} \left[ F_{kl} j^l + G_{kl} k^l \right]
\end{aligned}$$

using here the following formula :

(35)

$$F_{kl}\partial_i F^{li} + G_{kl}\partial_i G^{li} = (\frac{1}{2}\varepsilon_{klrs}G^{rs})\partial_i(\frac{1}{2}\varepsilon^{lipq}G_{pq}) + (-\frac{1}{2}\varepsilon_{klrs}F^{rs})\partial_i(-\frac{1}{2}\varepsilon^{lipq}F_{pq})$$
$$= \frac{1}{4}\left[\delta^{pqi}_{rsk}G^{rs}\partial_i G_{pq}\right] + \frac{1}{4}\left[\delta^{pqi}_{rsk}F^{rs}\partial_i F_{pq}\right]$$
$$= \left[G^{li}\partial_i G_{kl} + \frac{1}{4}\partial_k G^{pq}G_{pq}\right] + \left[F^{li}\partial_i F_{kl} \frac{1}{4}\partial_k F^{pq}F_{pq}\right]$$
$$= G^{li}\partial_i G_{kl} + F^{li}\partial_i F_{kl}$$

where : $F^{pq}F_{pq} = -G^{pq}G_{pq}$

Let us now write the symmetrized Lorentz force density by substituting (34) in (33) :

(36) $$f_k = \mu_0 \frac{du_k}{dt_0} = \frac{1}{c}\left[F_{kl}j^l + G_{kl}k^l\right]$$

Making use of definition (33), one can say that the Lorentz force can also be directly obtained from Maxwell's equations with the definition : $f_k = -\partial_i T^{(f)i}{}_k = \frac{1}{4\pi}[F_{kl}(14) + G_{kl}(15)]$.

The force acting on a charge particle with electric and magnetic charge, $q_e$ and $q_m$, (call dual-charged particles) and with a proper mass $m_0$, moving with velocity $u^k$, in the presence of fields is given by the Lorentz force law :

(37) $$F_k = m_0 \frac{du_k}{dt_0} = \frac{1}{c}\left[q_e F_{kl}u^l + q_m G_{kl}u^l\right]$$

The four-vector force is given by : $F_k = (\frac{\gamma}{c}\mathbf{F}\cdot\mathbf{v}, -\gamma\mathbf{F})$. With (7) and (8), the reader can check that the vectorial Lorentz force can be expressed as :

(38) $$\mathbf{F} = q_e(\mathbf{E} + \frac{\mathbf{v}}{c}\times\mathbf{B}) + q_m(\mathbf{B} - \frac{\mathbf{v}}{c}\times\mathbf{E})$$

The zero component of $F_k$ : $F_0 = \frac{\gamma}{c}\mathbf{F}\cdot\mathbf{v}$ is compatible with this result.

The equation (38) exhibits a symmetry not present in conventional electrodynamics. The magnetic monopoles are accelerated by a magnetic field and their paths are bent by an electric field. This result can be useful for monopoles detection systems.

## 5. Conclusion

The formulation of unified classical electromagnetism theory, associated with electric and magnetic sources $j^i$ and $k^i$, is developed here in a classical model. It is shown that it is possible to write, with a two potentials approach, an action integral from which one can derive, by least action principle, the symmetrized Maxwell's equations, as well as the equation of motion of dual-charged particles. A new original way is explored here for the derivation of the symmetrized Lorentz force by employing the energy-mommentum tensor conservation in curvilinear coordinates.